\documentclass{article}
\usepackage{spconf,amsmath,amssymb,graphicx,hyperref}

\usepackage{orcidlink}

\title{Rel2Traj: Inter-Source Relation-Aware Trajectory Generation for Text-To-Spatial-Audio}
\name{\begin{tabular}{c}
Yihang Bao$^{1}$\,\orcidlink{0009-0009-0809-4041},
Lingfeng Yao$^{2}$\,\orcidlink{0009-0006-2451-5633},
Shengyao Wang$^{1}$\,\orcidlink{0009-0004-9167-4234},
Xincong Zhong$^{1}$\,\orcidlink{0009-0009-0964-8827},\\
Junpei Xue$^{1}$\,\orcidlink{0009-0008-7165-9634},
Miao Pan$^{2}$\,\orcidlink{0000-0003-2138-4413},
Jiang Liu$^{1}$\,\orcidlink{0000-0003-0613-0298}
\end{tabular}}

\address{$^{1}$Waseda University, Tokyo, Japan \qquad
$^{2}$University of Houston, Houston, USA}

\begin{document}
%\ninept
%
\maketitle

\begin{abstract}
``A car drives from front left to front right while a dog follows it.'' Such a description specifies no particular trajectory, yet imposes a constraint on the joint motion of both sources that must hold at every frame. Existing text-to-spatial-audio methods can position and move each sound source according to its description, but none of them ensures or verifies that such inter-source relations hold. We therefore propose Rel2Traj, which parses the description into sources, relations, and anchors, jointly generates the trajectories of all sources using a relation-conditioned diffusion model, and finally renders sounds from a pretrained text-to-audio model along these trajectories to produce binaural audio. For training and evaluation, we construct a synthetic dataset from a controlled language of source motions and inter-source relations. A single frame-wise relational geometry defines both our evaluation metric, the Relation Satisfaction Rate (RSR), and an auxiliary training loss. On motion-relation combinations held out from training, Rel2Traj maintains all relation types, whereas every baseline fails on at least one. A listening test further confirms that listeners recognize the described relations. To our knowledge, this is the first text-to-spatial-audio work to model relations among moving sources as explicit geometric constraints.
\end{abstract}
\section{Introduction}
\label{sec:intro}

In recent years, rapid advances in generative audio models and multimodal learning have led to the development of spatial audio generation methods conditioned on different modalities. Visually guided methods generate or spatialize spatial audio from 360° videos or images~\cite{morgado360,visualsound25d,visage}. Existing text-driven methods instead extract spatial information from descriptions of a single source’s static position or motion~\cite{immersediffusion,sonicmotion}, or the absolute directions of individual sources in a multi-source scene~\cite{bewo}. However, a sound scene often contains several interrelated sources moving simultaneously. Descriptions of such scenes mainly specify how each source moves relative to the listener and the spatial relations among the sources. Such a relation constrains the joint motion of multiple sources at each frame while allowing many valid combinations of trajectories. A generator must therefore produce valid sets of trajectories rather than a single trajectory. It must also be possible to verify at each frame whether the generated trajectories satisfy the relation.

Existing text-to-spatial-audio methods fall into two families, and both give the spatial information as the absolute direction of each source. The first is end-to-end generation, which uses a diffusion model to directly produce spatial audio with a static source~\cite{immersediffusion,dualspec} or with a source moving in a straight line between two points~\cite{bewo,sonicmotion}. In these methods, spatial properties are encoded only in the generated audio and can only be assessed after generation. Such assessment is unreliable for multiple or moving sources ~\cite{bewo}. The second is the cascaded approach, which parses the text into an explicit spatial representation for each source, then generates and renders each sound separately. TTMBA~\cite{ttmba} and STASE~\cite{stase} parse static coordinates, while Text2Move~\cite{text2move} generates a trajectory, although only for a single source. Positions are thus explicit in cascaded methods, but the relations that sources should maintain with one another are reflected neither in the text nor in the pipeline. Such relations have so far appeared only in auditory understanding~\cite{bat} and in monaural spatialization~\cite{tas_inthewild}, and only for static sources. None of these methods therefore models the relations among moving sources. 
\begin{figure*}[t]
  \centering
  \includegraphics[width=\textwidth]{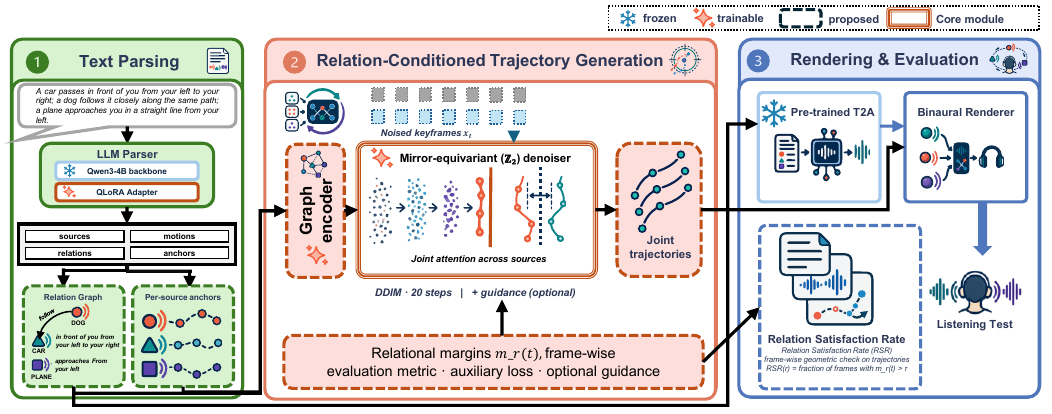}
  \caption{Rel2Traj: (1)~text parsing into a relation graph and per-source anchors, (2)~joint trajectory generation by a relation-conditioned keyframe diffusion model, and (3)~binaural rendering and evaluation. The frame-wise relational margins (bottom) define the RSR metric and also serve as an auxiliary loss and optional guidance.}
  \label{fig:framework}
\end{figure*}

We propose Rel2Traj, a framework for multi-source spatial audio generation conditioned on inter-source relations. Our contributions are threefold. (i) Relation-aware joint trajectory generation. We fine-tune a large language model to parse the description into sources, relations, and per-source anchors. A diffusion model conditioned on the resulting relation graph and anchors generates all trajectories at once, with cross-source attention coordinating them. (ii) Differentiable and verifiable relation geometry. We express each relation as a per-frame signed margin that defines our evaluation metric, the Relation Satisfaction Rate (RSR), and doubles as an auxiliary training loss. (iii) A controlled dataset with a compositional split and dual validation. We synthesize a binaural dataset from a controlled language of source motions and relations, hold out motion--relation combinations from training, and validate the results both geometrically and perceptually. To the best of our knowledge, this is the first work in text-to-spatial-audio generation that models inter-source relations as explicit geometric constraints.

\section{Method}
\label{sec:format}

\subsection{Overview}
\label{sec:overview}

Rel2Traj integrates three components (Fig.~\ref{fig:framework}): a QLoRA-fine-tuned LLM parser that converts the description into sources with motion types, spatial relations, and per-source anchors; a relation-conditioned keyframe diffusion model that generates the trajectories of all sources jointly under these conditions; and a physics-based binaural renderer that renders per-class sounds from a pretrained text-to-audio model along the trajectories. 

\subsection{Parser and Dual-Level Conditioning}
\label{sec:conditioning}

We fine-tune Qwen3-4B with QLoRA ($r{=}16$, $\alpha{=}32$) on the description--structure pairs of our dataset. The parser converts a description into three fields: the list of sources, each with a motion type; the list of relations, each with a type, a subject, a reference (another source or the listener), and an optional time scope; and ten discrete anchor fields per source describing its coarse placement and path.

We represent each scene as a directed graph $G{=}(V,E)$ with one node per source and one typed edge per inter-source relation. The initial node state sums the embeddings of the motion type, the anchor fields, and the listener-relative relations. A graph encoder then performs two rounds of typed message passing over $G$ with a gated recurrent update in the style of GGNN~\cite{ggnn}, yielding one condition vector per source. These dual-level conditions are the only conditioning input to the diffusion model: the relation graph constrains sources with respect to one another, and the anchors constrain each source individually.

\subsection{Relation-Conditioned Keyframe Diffusion}
\label{sec:diffusion}

\textbf{Trajectory representation.}
Each source trajectory is represented by $K{=}16$ Cartesian keyframes. The $N$ sources of a scene form one sample of shape $N{\times}K{\times}3$, and the diffusion process acts on the whole sample. Keyframes are interpolated to 30~fps after generation.

\textbf{Denoiser.}
The denoiser follows the Transformer diffusion models used in motion generation~\cite{mdm,motiondiffuse}. To this design we add a mirror-equivariant structure and joint attention across sources. The network is a $\mathbb{Z}_2$ mirror-equivariant Transformer with width 192, 4 layers, and 8 heads. It flattens the $N{\times}K$ keyframes into one token sequence and attends over all of them, so the trajectories are coordinated at every denoising step. We use the mirror group because left--right chirality is the central symmetry of spatial language, whereas front--back and up--down are not symmetric in listener-centered semantics. The per-source condition vectors and a timestep embedding are added to every keyframe token. The graph encoder is mirror-equivariant as well, so the condition vectors transform consistently and the whole generator commutes with the mirror by construction. The generator has 1.52M parameters in total, 0.92M of them in the denoiser.

\textbf{Training and sampling.}
With a cosine noise schedule ($T{=}100$) the network predicts the noise $\varepsilon$, and the loss is
\begin{equation}
\mathcal{L} = \mathbb{E}_{t,\varepsilon}\Big[\,\|\hat{\varepsilon}-\varepsilon\|_2^2
+ \lambda_{\mathrm{geo}}\,\bar{\alpha}_t\big(\mathcal{L}_{\mathrm{rel}}(\hat{x}_0) + 0.1\,\mathcal{L}_{\mathrm{smooth}}(\hat{x}_0)\big)\Big],
\label{eq:loss}
\end{equation}
with $\lambda_{\mathrm{geo}}{=}0.1$, where $\hat{x}_0$ is the clean estimate recovered from the current prediction, $\mathcal{L}_{\mathrm{rel}}$ is a squared hinge on the relational margins of $\hat{x}_0$ defined below, $\mathcal{L}_{\mathrm{smooth}}$ penalizes second differences, and $\bar{\alpha}_t$ down-weights the geometric terms at high noise levels, at which $\hat{x}_0$ is unreliable. Sampling uses DDIM with 20 steps. The geometric term is auxiliary: removing $\mathcal{L}_{\mathrm{rel}}$ from training leaves RSR unchanged (Table~\ref{tab:ablation}). Margin-gradient guidance on $\hat{x}_0$~\cite{gmd,omnicontrol} is optional and disabled for all reported results.

\subsection{Dataset Construction}
\label{sec:dataset}

Existing datasets lack descriptions of relations among moving sources paired with ground-truth trajectories, so we construct a synthetic one from a controlled language that generates both. The language defines 15 motion primitives and 25 relation words (8 listener sectors, 10 listener-relative, and 7 inter-source relations), plus 10 discrete anchor fields per source. All spatial semantics are listener-centered rather than world-axis based, following the controlled-language tradition of CLEVR~\cite{clevr} and spatial-language research~\cite{levinson2003}. 

The dataset has 5091 training, 242 validation, and 1020 compositional test scenes. A training scene contains one to three sources, with two or three sources in 82\% of the scenes, and one to ten relations. Scenes last 8--14~s. Ground-truth trajectories record the azimuth, elevation, and distance of every source at 30~fps. About a third of the relations are restricted to a time scope, and the rest hold over the whole scene. The compositional split is pre-registered. Five motion--relation cells were frozen before any training and appear only in the test set, while both components of every cell occur in training but never together. The test therefore probes unseen combinations rather than unseen relation types.

Sound material comes from ESC-50 recordings~\cite{esc50} and from clips generated offline per class by a pretrained text-to-audio model and automatically screened~\cite{stableaudioopen}. The binaural renderer realizes Doppler shift, HRTF interpolation, and room reflections~\cite{allen1979}, normalizes loudness with gated LUFS~\cite{bs1770}, and has no learnable parameters.

\subsection{The RSR Metric}
\label{sec:rsr}

For each relation type, we define a signed margin $m_r(t)$ at each frame and consider the relation satisfied when the margin exceeds a small dead-zone threshold $\tau$. For \emph{closer\_than}, the margin is the difference between the reference's and the subject's distance to the listener,
\begin{equation}
m_{\mathrm{closer}}(t) = d_{\mathrm{ref}}(t) - d_{\mathrm{subj}}(t);
\label{eq:margin}
\end{equation}
relations with several conditions, such as \emph{follow}, take the minimum of their margins. The Relation Satisfaction Rate (RSR) is the fraction of in-scope frames whose margin exceeds the dead zone,
\begin{equation}
\mathrm{RSR}(r) = \frac{1}{|S_r|}\sum_{t \in S_r} \mathbf{1}\big[\, m_r(t) > \tau \,\big], \qquad \tau = 0.05,
\label{eq:rsr}
\end{equation}
where $S_r$ contains the frames within the relation's time scope and, for motion relations, only those in which the source moves. For positional relations the margin and $\tau$ are in meters. For motion relations the margin is a rate in m/s, except for \emph{same\_direction}, \emph{opposite\_direction}, and \emph{follow}, whose margins take the minimum of a heading cosine and pair distances in meters.

The evaluated outputs are never optimized against the metric directly, because training uses the margins on training scenes only, guidance is off in all reported results, and the test set consists of held-out combinations. Ground truth is built from motion primitives with a 95\% quality gate, which is why it scores 0.996 rather than 1.0.

\section{Results and discussions}
\label{sec:pagestyle}

\subsection{Setup and Baselines}
\label{sec:setup}

All systems are evaluated on two frozen splits: an in-distribution validation set (242 scenes) and a compositional test set containing only the held-out combinations (1020 scenes). The metric is RSR. The scene score is averaged over ten samples per scene and then over scenes, with no selection. Per-type numbers average over all instances of that type across the ten samples. We additionally report jerk, the mean magnitude of the third derivative of position, and one-to-many diversity, the mean pairwise displacement, in meters, between the ten samples of a scene.

We compare against a ground-truth reference and three baselines. Ground-truth trajectories give the upper bound of the metric. A rule-based solver takes the motion types and the explicit relations as input and adjusts one trajectory set by minimizing hinge penalties on the relational margins. LLM-direct prompts Qwen3-8B to output trajectory coordinates directly from the description, without any geometric modeling. The independent per-source baseline shares the architecture, parameter count, and training budget of Rel2Traj, but removes the cross-source information flow and masks the anchors, so that each source is generated on its own. This corresponds to the per-source decomposition of earlier single-source trajectory prediction~\cite{text2move}. Except for LLM-direct, which reads the raw text, all systems in this section are conditioned on the ground-truth structure of each scene, so that the tables isolate generation from parsing. The fine-tuned parser is evaluated separately on the 242 validation scenes, where it recovers every relation and motion type exactly and predicts the ten anchor fields with 85--100\% accuracy.

\begin{table}[t]
\centering
\small
\caption{RSR (scene mean over ten samples, no selection) on the validation (val) and compositional test (comp) splits. Jerk and diversity are reported on comp. The solver and LLM-direct are deterministic and yield one trajectory set per scene.}
\label{tab:main}
\begin{tabular*}{\columnwidth}{@{\extracolsep{\fill}}lcccc@{}}
\hline
 & \multicolumn{2}{c}{RSR} & Jerk & Div. \\
System & val & comp & $\downarrow$ & (m) \\
\hline
Ground truth (ref.) & 0.996 & 0.995 & 0.5 & -- \\
Rule-based solver & 0.963 & 0.919 & 8.0 & single \\
LLM-direct & 0.610 & 0.530 & 9.7 & single \\
Indep.\ per-source & 0.896 & 0.869 & 9.2 & 1.56 \\
Rel2Traj & \textbf{0.987} & \textbf{0.983} & 9.8 & 0.44 \\
\hline
\end{tabular*}
\end{table}

\begin{table}[t]
\centering
\small
\caption{Ablation of the relational mechanisms (comp split unless noted).}
\label{tab:ablation}
\begin{tabular*}{\columnwidth}{@{\extracolsep{\fill}}lccccc@{}}
\hline
 & \multicolumn{2}{c}{RSR} & & & Div. \\
Variant & val & comp & \emph{follow} & \emph{behind} & (m) \\
\hline
Rel2Traj (full) & 0.987 & 0.983 & 0.992 & 0.780 & 0.44 \\
w/o $\mathcal{L}_{\mathrm{rel}}$ & 0.988 & 0.987 & 0.991 & 0.923 & 0.45 \\
w/o coupling & 0.983 & 0.969 & 0.692 & 0.929 & 0.45 \\
w/o anchors & 0.952 & 0.938 & 0.999 & 0.053 & 1.30 \\
\hline
\end{tabular*}
\end{table}

\subsection{Geometric Evaluation}
\label{sec:geo}

Table~\ref{tab:main} compares Rel2Traj with the reference and the three baselines. Rel2Traj outperforms all three baselines in RSR on both splits, achieving 0.983 on the held-out combinations. The rule-based solver is close behind but fails on same- and opposite-direction relations, scoring at most 0.52 on either split, as its local optimizer does not reach a satisfying layout. The independent per-source baseline collapses on \emph{follow} and \emph{behind}. Two individually plausible trajectories do not satisfy the relation once combined. Every baseline thus fails on at least one relation type. For Rel2Traj no type collapses, and its lowest cell is \emph{behind} on the held-out combinations (0.780). Rel2Traj has the highest jerk in Table~\ref{tab:main}, so its advantage lies in coordination rather than smoothness. The independent baseline shows the largest diversity, but this spread reflects uncoordinated sources rather than more valid solutions.

Table~\ref{tab:ablation} separates the roles of the three relational mechanisms. Removing cross-source coupling mainly degrades \emph{follow} (0.992 to 0.692), while removing anchors collapses \emph{behind} (0.780 to 0.053) and triples diversity (0.44 to 1.30~m), indicating that coupling handles fine coordination and anchors provide coarse placement. Removing both yields the independent baseline of Table~\ref{tab:main}. Removing the explicit relational loss leaves compositional RSR unchanged (0.983 against 0.987, with 22 of 25 relation types within 0.01), showing that relational consistency is learned through structured conditioning and joint generation rather than through direct optimization of the evaluation geometry. The mirror-equivariant architecture enables generalization to the two held-out chirality cells, \emph{counterclockwise} and \emph{follow}. A non-equivariant Transformer of the same width and depth reaches the same validation RSR (0.988) but only 0.835 on the compositional split, and training it additionally with a mirror-consistency loss does not help (0.720 against 0.727 on the 579 scenes of the two cells), whereas the equivariant model, with a third fewer parameters, reaches 0.993 on the same scenes.

\subsection{Perceptual Evaluation}
\label{sec:perc}

We ran a remote, anonymous listening test with 24 valid participants (2 excluded by pre-registered rules). In each item, participants heard a rendered two-source scene and chose which of two descriptions, differing only in the inter-source relation, matched what they heard. The 28 items covered six relation types, with ground-truth and Rel2Traj renderings mixed and one of three HRTF sets assigned, both blind to the participant. The first pre-registered criterion asks whether the relations are audible at all: recognition of Rel2Traj relations reached 86.0\% (Table~\ref{tab:listening}), above chance. The second asks how far the generations are from ground truth: the gap was $-3.3$ points, within the preset $\pm$10, and an equivalence test (TOST) confirmed it. Four additional ABX items tested whether ground-truth and generated renderings could be told apart at all, and discrimination was at chance. These results are limited to the pre-registered relation types and show that the generated scenes are as valid a realization of the description as the ground truth, not that they reproduce it. In a few items the relation held geometrically but only by a small margin, for example two sources at nearly the same distance from the listener, and participants could not hear it. 

\begin{table}[t]
\centering
\small
\caption{Pre-registered listening-test criteria ($n=24$).}
\label{tab:listening}
\begin{tabular*}{\columnwidth}{@{\extracolsep{\fill}}llc@{}}
\hline
Criterion & Test & Result \\
\hline
Rel2Traj $>$ chance & $t$-test & \textbf{86.0\%}, $p<10^{-15}$ \\
$|\mathrm{GT}-\mathrm{Rel2Traj}| \le 10$ pts & TOST & $-3.3$ pts, $p<0.05$ \\
ABX $\approx$ chance & binomial & 46.9\%, $p=0.61$ \\
\hline
\end{tabular*}
\end{table}

\section{Conclusion}
\label{sec:conclusion}

Rel2Traj jointly generates trajectories conditioned on explicit inter-source relations and renders binaural audio, using frame-wise relational margins to verify relation satisfaction. On held-out motion--relation combinations, it maintains all relation types, and listeners recognize the described relations. Limitations include the controlled language, at most one inter-source relation per scene, run-to-run variation on held-out \emph{behind}, and the exclusion of purely vertical motions. Audio examples and an interactive trajectory viewer are available at \url{https://echo-sculpture.github.io/Rel2TrajSA/}.

\bibliographystyle{IEEEbib}
\bibliography{strings,refs}

\begin{thebibliography}{10}

\bibitem{morgado360}
Pedro Morgado, Nuno Vasconcelos, Timothy Langlois, and Oliver Wang,
\newblock ``Self-supervised generation of spatial audio for {360$^\circ$} video,''
\newblock in {\em Advances in Neural Information Processing Systems}, 2018, vol.~31.

\bibitem{visualsound25d}
Ruohan Gao and Kristen Grauman,
\newblock ``{2.5D} visual sound,''
\newblock in {\em Proc. CVPR}, 2019, pp. 324--333.

\bibitem{visage}
Jaeyeon Kim, Heeseung Yun, and Gunhee Kim,
\newblock ``{ViSAGe}: Video-to-spatial audio generation,''
\newblock in {\em Proc. ICLR}, 2025.

\bibitem{immersediffusion}
Mojtaba Heydari, Mehrez Souden, Bruno Conejo, and Joshua Atkins,
\newblock ``{ImmerseDiffusion}: A generative spatial audio latent diffusion model,''
\newblock in {\em Proc. IEEE ICASSP}, 2025, pp. 1--5.

\bibitem{sonicmotion}
Christian Templin, Yanda Zhu, and Hao Wang,
\newblock ``Generating moving {3D} soundscapes with latent diffusion models,''
\newblock in {\em Proc. IEEE ICASSP}, 2026, pp. 2856--2860.

\bibitem{bewo}
Peiwen Sun, Sitong Cheng, Xiangtai Li, Zhen Ye, Huadai Liu, Honggang Zhang, Wei Xue, and Yike Guo,
\newblock ``Both ears wide open: Towards language-driven spatial audio generation,''
\newblock in {\em Proc. ICLR}, 2025.

\bibitem{dualspec}
Lei Zhao, Sizhou Chen, Linfeng Feng, Jichao Zhang, Xiao-Lei Zhang, Chi Zhang, and Xuelong Li,
\newblock ``{DualSpec}: Text-to-spatial-audio generation via dual-spectrogram guided diffusion model,''
\newblock {\em IEEE Trans. Multimedia}, vol. 28, pp. 7112--7123, 2026.

\bibitem{ttmba}
Yuxuan He, Xiaoran Yang, Ningning Pan, and Gongping Huang,
\newblock ``{TTMBA}: Towards text to multiple sources binaural audio generation,''
\newblock in {\em Proc. Interspeech}, 2025, pp. 4228--4232.

\bibitem{stase}
Tutti Chi, Letian Gao, and Yixiao Zhang,
\newblock ``{STASE}: A spatialized text-to-audio synthesis engine for music generation,'' Presented at the LLM4Music Workshop at ISMIR, 2025.

\bibitem{text2move}
Yunyi Liu, Shaofan Yang, Kai Li, and Xu~Li,
\newblock ``{Text2Move}: Text-to-moving sound generation via trajectory prediction and temporal alignment,''
\newblock in {\em Proc. IEEE ICASSP}, 2026, pp. 15457--15461.

\bibitem{bat}
Zhisheng Zheng, Puyuan Peng, Ziyang Ma, Xie Chen, Eunsol Choi, and David Harwath,
\newblock ``{BAT}: Learning to reason about spatial sounds with large language models,''
\newblock in {\em Proc. ICML}, 2024, pp. 61454--61469.

\bibitem{tas_inthewild}
Tianrui Pan, Jie Liu, Zewen Huang, Jie Tang, and Gangshan Wu,
\newblock ``In-the-wild audio spatialization with flexible text-guided localization,''
\newblock in {\em Proc. ACL}, 2025, pp. 1989--2001.

\bibitem{ggnn}
Yujia Li, Daniel Tarlow, Marc Brockschmidt, and Richard Zemel,
\newblock ``Gated graph sequence neural networks,''
\newblock in {\em Proc. ICLR}, 2016.

\bibitem{mdm}
Guy Tevet, Sigal Raab, Brian Gordon, Yonatan Shafir, Daniel Cohen-Or, and Amit~H. Bermano,
\newblock ``Human motion diffusion model,''
\newblock in {\em Proc. ICLR}, 2023.

\bibitem{motiondiffuse}
Mingyuan Zhang, Zhongang Cai, Liang Pan, Fangzhou Hong, Xinying Guo, Lei Yang, and Ziwei Liu,
\newblock ``{MotionDiffuse}: Text-driven human motion generation with diffusion model,''
\newblock {\em IEEE Trans. Pattern Anal. Mach. Intell.}, vol. 46, no. 6, pp. 4115--4128, 2024.

\bibitem{gmd}
Korrawe Karunratanakul, Konpat Preechakul, Supasorn Suwajanakorn, and Siyu Tang,
\newblock ``Guided motion diffusion for controllable human motion synthesis,''
\newblock in {\em Proc. ICCV}, 2023, pp. 2151--2162.

\bibitem{omnicontrol}
Yiming Xie, Varun Jampani, Lei Zhong, Deqing Sun, and Huaizu Jiang,
\newblock ``{OmniControl}: Control any joint at any time for human motion generation,''
\newblock in {\em Proc. ICLR}, 2024.

\bibitem{clevr}
Justin Johnson, Bharath Hariharan, Laurens van~der Maaten, Li~Fei-Fei, C.~Lawrence Zitnick, and Ross Girshick,
\newblock ``{CLEVR}: A diagnostic dataset for compositional language and elementary visual reasoning,''
\newblock in {\em Proc. CVPR}, 2017, pp. 1988--1997.

\bibitem{levinson2003}
Stephen~C. Levinson,
\newblock {\em Space in Language and Cognition: Explorations in Cognitive Diversity},
\newblock Cambridge University Press, Cambridge, UK, 2003.

\bibitem{esc50}
Karol~J. Piczak,
\newblock ``{ESC}: Dataset for environmental sound classification,''
\newblock in {\em Proc. ACM Multimedia}, 2015, pp. 1015--1018.

\bibitem{stableaudioopen}
Zach Evans, Julian~D. Parker, CJ~Carr, Zack Zukowski, Josiah Taylor, and Jordi Pons,
\newblock ``{Stable Audio Open},''
\newblock in {\em Proc. IEEE ICASSP}, 2025, pp. 1--5.

\bibitem{allen1979}
Jont~B. Allen and David~A. Berkley,
\newblock ``Image method for efficiently simulating small-room acoustics,''
\newblock {\em J. Acoust. Soc. Am.}, vol. 65, no. 4, pp. 943--950, 1979.

\bibitem{bs1770}
{ITU-R},
\newblock ``Recommendation {BS.1770-4}: Algorithms to measure audio programme loudness and true-peak audio level,'' International Telecommunication Union, 2015.

\end{thebibliography}

\end{document}